\documentclass{iau}
\usepackage{graphicx}
\usepackage{amsmath}
\usepackage{url}

\title[Solar neighbourhood age scale] %% give here short title %%
{Isochronal age scale of young moving groups in the solar neighbourhood}

\author[C.~P.~M.~Bell et al.]   %% give here short author list %%
{Cameron P. M. Bell$^1$, Eric E. Mamajek$^1$ \and Tim Naylor$^2$}%, N. J. Mayne$^2$, R. D. Jeffries$^3$ \and S. P. Littlefair$^4$}

\affiliation{$^1$Department of Physics \& Astronomy, University of Rochester, Rochester, NY 14627, USA\\email: {\tt cbell@pas.rochester.edu}\\
$^2$School of Physics and Astronomy, University of Exeter, Exeter EX4 4QL, UK}
\pubyear{2015}
\volume{314}  %% insert here IAU Symposium No.
\pagerange{119--126}
\jname{Young Stars \& Planets Near the Sun}
\editors{J. H. Kastner, B. Stelzer, \& S. A. Metchev, eds.}
\begin{document}

\maketitle

\begin{abstract}
We present a self-consistent, absolute isochronal age scale for young $(\lesssim 200\,\rm{Myr})$, nearby $(\lesssim 100\,\rm{pc})$ moving groups, which is consistent with recent lithium depletion boundary ages for both the $\beta$ Pic and Tucana-Horologium moving groups. This age scale was derived using a set of semi-empirical pre-main-sequence model isochrones that incorporate an empirical colour-$T_{\rm{eff}}$ relation and bolometric corrections based on the observed colours of Pleiades members, with theoretical corrections for the dependence on log\,$g$.
Absolute ages for young, nearby groups are vital as these regions play a crucial role in our understanding of the early evolution of low- and intermediate-mass stars, as well as providing ideal targets for direct imaging and other measurements of dusty debris discs, substellar objects and, of course, extrasolar planets.
\end{abstract}

\firstsection % if your document starts with a section,
\section{Introduction}
\label{introduction}

Over the past couple of decades several hundred low- and intermediate-mass stars have been identified within $\simeq 100\,\rm{pc}$ of the Sun as part of a concerted effort to identify and characterise the young solar neighbourhood. The distribution of these stars on the sky is not uniform, instead they comprise dispersed, predominantly unbound associations in which the members share a common space motion. Although the age ordering of these young associations is well-constrained i.e. the AB Dor moving group is older than the Tucana-Horologium moving group (Tuc-Hor), which in turn is older than the $\beta$ Pic moving group (BPMG), the absolute ages of these groups are still under-constrained.

Theoretical model isochrones are (arguably) the most commonly used method of age-dating young (presumably coeval) stellar populations. Although the use of maximum-likelihood fitting techniques to subjectively fit model isochrones has been used for more distant Galactic clusters (e.g. NGC~2547 in \cite[Naylor \& Jeffries 2006]{Naylor2006}), such methods have not yet been applied to the young groups within $\simeq 100\,\rm{pc}$. Additionally, for these young groups there has been a distinct lack of homogeneity when it comes to the fitting of model isochrones, particularly in terms of the models adopted and the photometric bandpasses used to construct the colour-magnitude diagrams (CMDs). Furthermore, recent lithium depletion boundary (LDB) ages, which are advocated in the recent review of young stellar ages by \cite[Soderblom et al. (2014)]{Soderblom2014} to provide our best chance of establishing a \emph{reliable} and \emph{robust} age scale in the range $20-200\,\rm{Myr}$, have been demonstrated to be systematically older when compared to isochronal ages (see e.g. \cite[Kraus et al. 2014]{Kraus2014}).

In this contribution we provide a self-consistent, absolute isochronal age scale for young ($\lesssim 200\,\rm{Myr}$) moving groups within 100\,pc of the Sun. This age scale is based on homogeneous fitting of photometric data in the $M_{V}, V-J$ CMD using a maximum-likelihood fitting statistic in conjunction with semi-empirical model isochrones. We find that the isochronal ages for both the BPMG and Tuc-Hor are consistent with recent LDB ages, thereby instilling confidence in our isochronal age scale.

\section{Sample of young, nearby moving groups}
\label{clusters_and_associations}

In this contribution we focus on the young, nearby moving groups within 100\,pc; specifically the AB Dor moving group, Argus association, BPMG, Carina association, Columba association, $\eta$ Cha cluster, Tuc-Hor, TW Hya association (TWA), and 32 Ori group (see also the recent review by \cite[Torres et al. 2008]{Torres2008}). For each group, we have assembled a list of members and candidate members (hereafter simply referred to together as `members') from the literature. For the inclusion of candidate members, we require that they have a membership probability of $\geq 90\%$ as calculated using the Bayesian Analysis for Nearby Young AssociatioNs (BANYAN; see e.g. \cite[Malo et al. 2013]{Malo2013}). Furthermore, we require that any candidate members have a measured radial velocity and/or trigonometric parallax which are/is consistent with membership for a given young group.

Our sample of members (not accounting for unresolved multiples) includes 89 members of the AB Dor moving group, 27 members of Argus, 97 members of the BPMG, 12 members of Carina, 50 members of Columba, 18 members of $\eta$ Cha, 189 members of Tuc-Hor, 30 members of TWA, and 14 members of 32 Ori.

Whilst all of our members have counterparts in the Two-Micron All-Sky Survey Point Source Catalog (2MASS PSC; \cite[Cutri et al. 2003]{Cutri2003}), near-IR CMDs are not ideal for age determination primarily because for $2500 \lesssim T_{\rm{eff}} \lesssim 4000\,\rm{K}$ the loci of young clusters becomes vertical ($J-K_{\rm{s}} \simeq 0.9\,\rm{mag}$), and hence degenerate with age. Therefore we supplement our near-IR photometry with $V$-band data and derive ages for our sample of groups in the $M_{V}, V-J$ CMD so as to minimise the effects of circumstellar material on the $K_{\rm{s}}$ magnitudes for the youngest groups (especially $\eta$ Cha and TWA).

For assigning distances to each star in our list of members we prefer to adopt trigonometric parallax measurements, however when this is not an option (or where the parallax estimates appear to be in error or have unusually large uncertainties) we either adopt a kinematic distance from the literature or derive one using the `moving cluster' method (see e.g. \cite[Mamajek 2005]{Mamajek05}).

\section{Semi-empirical model isochrones}
\label{semi_empirical_model_isochrones}

\begin{figure}[t]
% \vspace*{-2.0 cm}
\centering
%  If you want to test that eps file inclusion works works, you can grab the eps figure file below from
% http://www.iau.org/static/scientific_meetings/authors/
% and then uncomment the following line.
\includegraphics[width=0.85\textwidth]{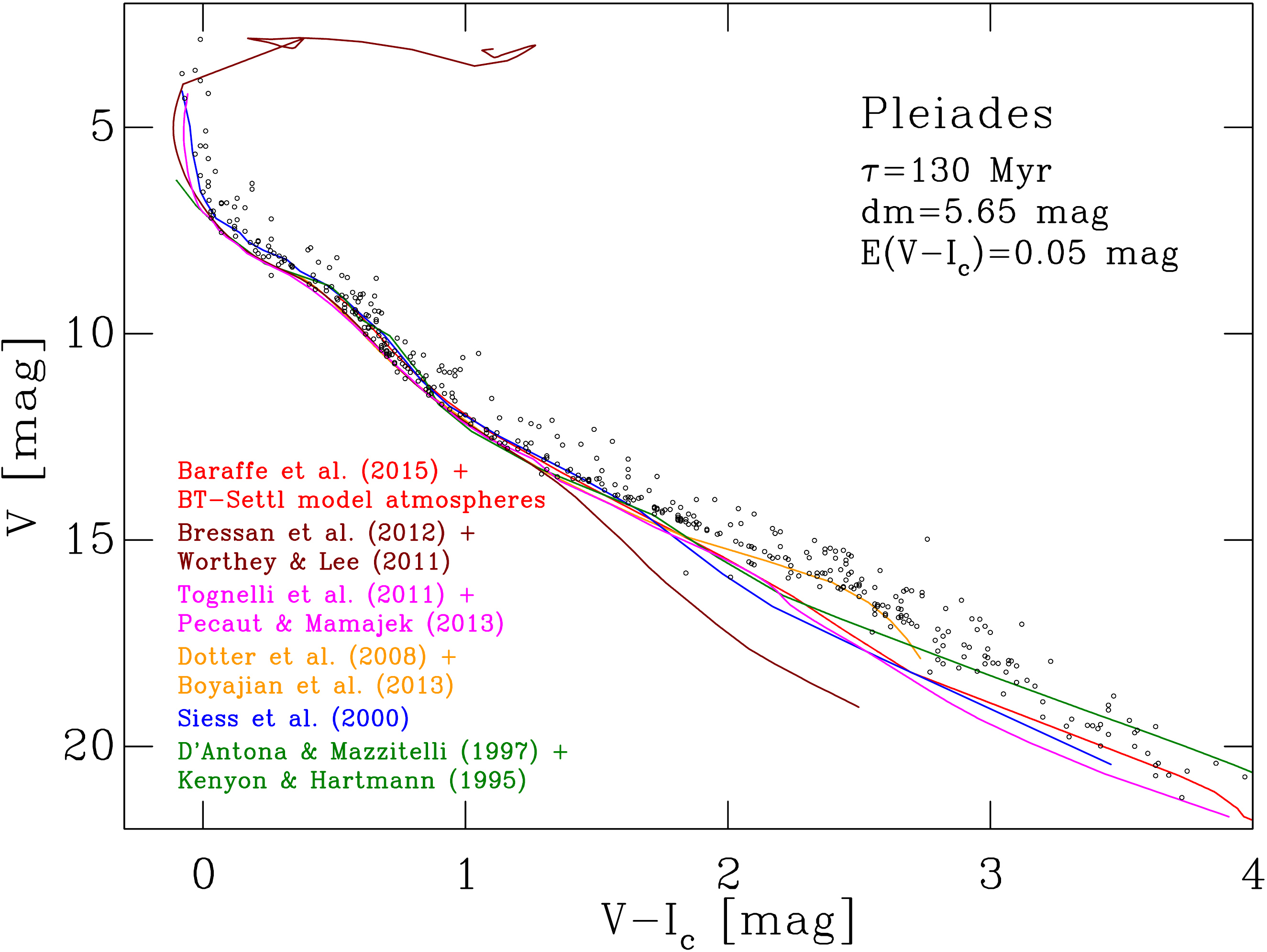} 
%\includegraphics[width=0.85\textwidth]{temp.pdf} 
% \vspace*{-1.0 cm}
 \caption{The $V, V-I_{\rm{c}}$ CMD of the Pleiades with several sets of model isochrones overlaid. The upper reference refers to the interior models, whereas the lower reference corresponds to the colour-$T_{\rm{eff}}$ relation and bolometric corrections adopted to transform the model into CMD space.}
   \label{fig:pleiades_cmd}
\end{figure}

There are large discrepancies between pre-main-sequence (pre-MS) tracks and observational data (such as dynamical masses, age trends, etc.; see e.g. \cite[Hillenbrand \& White 2004]{Hillenbrand2004}; Feiden this volume). In \cite[Bell et al. (2012)]{Bell2012} we discussed the reasons why the Pleiades represents an ideal benchmark cluster for the comparison of the models to the data, however chief among these is the fact that both the distance and age have been determined independently of the use of model isochrones. Fig.~\ref{fig:pleiades_cmd}, which shows the $V, V-I_{\rm{c}}$ CMD of the Pleiades with several commonly adopted sets of model isochrones overlaid, effectively demonstrates the aforementioned discrepancies. At $T_{\rm{eff}} \lesssim 4300\,\rm{K}$, the models overestimate the flux in the optical by up to a factor of two. Hence, if we are to derive consistent ages for young stellar populations using pre-MS model isochrones we must first perform some form of empirical correction to the models for $T_{\rm{eff}}$ lower than $\simeq 4300\,\rm{K}$.

In \cite[Bell et al. (2013, 2014)]{Bell2013,Bell2014)} we introduced a method of creating semi-empirical pre-MS model isochrones using the observed colours of young stars in the Pleiades, in addition to incorporating theoretical corrections for the dependence on the surface gravity (log\,$g$). Briefly, we used binary systems with well-constrained dynamical masses to demonstrate that the model predictions for the $K_{\rm{s}}$-band are essentially correct. This then allows us to use the $K_{\rm{s}}$-band magnitude along our fiducial locus as a $T_{\rm{eff}}$ indicator, but also ensures that our semi-empirical model isochrones have a reliable mass scale tied to that of the fiducial cluster. On the assumption that the $K_{\rm{s}}$-band magnitudes predicted by the models are correct, we can then use CMDs with the $K_{\rm{s}}$-band on both axes (e.g. $K_{\rm{s}}, V-K_{\rm{s}}$) to calculate any discrepancy between the fiducial locus and the model isochrone, which in this case would correspond to the requisite empirical correction in the $V$-band necessary to fit the Pleiades at a given age and distance. This empirical correction is then applied to the theoretical bolometric correction grid (calculated from atmospheric models and a function of $T_{\rm{eff}}$ and log$\,g$) at the appropriate $T_{\rm{eff}}$ irrespective of its log$\,g$. This process is then repeated for other photometric bandpasses to create sets of semi-empirical bolometric corrections. The semi-empirical pre-MS model isochrones used in this contribution are based on existing stellar interior models coupled with these newly derived semi-empirical bolometric corrections. In this contribution we adopt four sets of interior models, namely those of \cite[Dotter et al. (2008)]{Dotter2008}, \cite[Tognelli et al. (2011)]{Tognelli2011}, \cite[Bressan et al. (2012)]{Bressan2012}, and \cite[Baraffe et al. (2015]{Baraffe2015}; hereafter referred to as Dartmouth, Pisa, PARSEC, and BHAC15 respectively)\footnote[1]{A subset of these models is available via the Cluster Collaboration isochrone server \protect\url{http://www.astro.ex.ac.uk/people/timn/isochrones/}}.

\section{Isochronal age scale for young, nearby groups}
\label{isochronal_age_scale_for_young_clusters_and_associations}

We use the $\tau^{2}$ fitting statistic of \cite[Naylor \& Jeffries (2006)]{Naylor2006} and \cite[Naylor (2009)]{Naylor2009} to derive ages from the $M_{V}, V-J$ CMDs of our sample of young groups. The $\tau^{2}$ fitting statistic is particularly well-suited to this task because it allows for the effects of binarity, yields reliable uncertainties on the derived parameters and provides a goodness-of-fit test. We have made some modifications to the statistic since those of \cite[Naylor (2009)]{Naylor2009}, however for the purposes of this contribution we refrain from detailing these and instead refer the reader to Bell et al. (2015).

\begin{figure}[t]
\centering
\includegraphics[width=\textwidth]{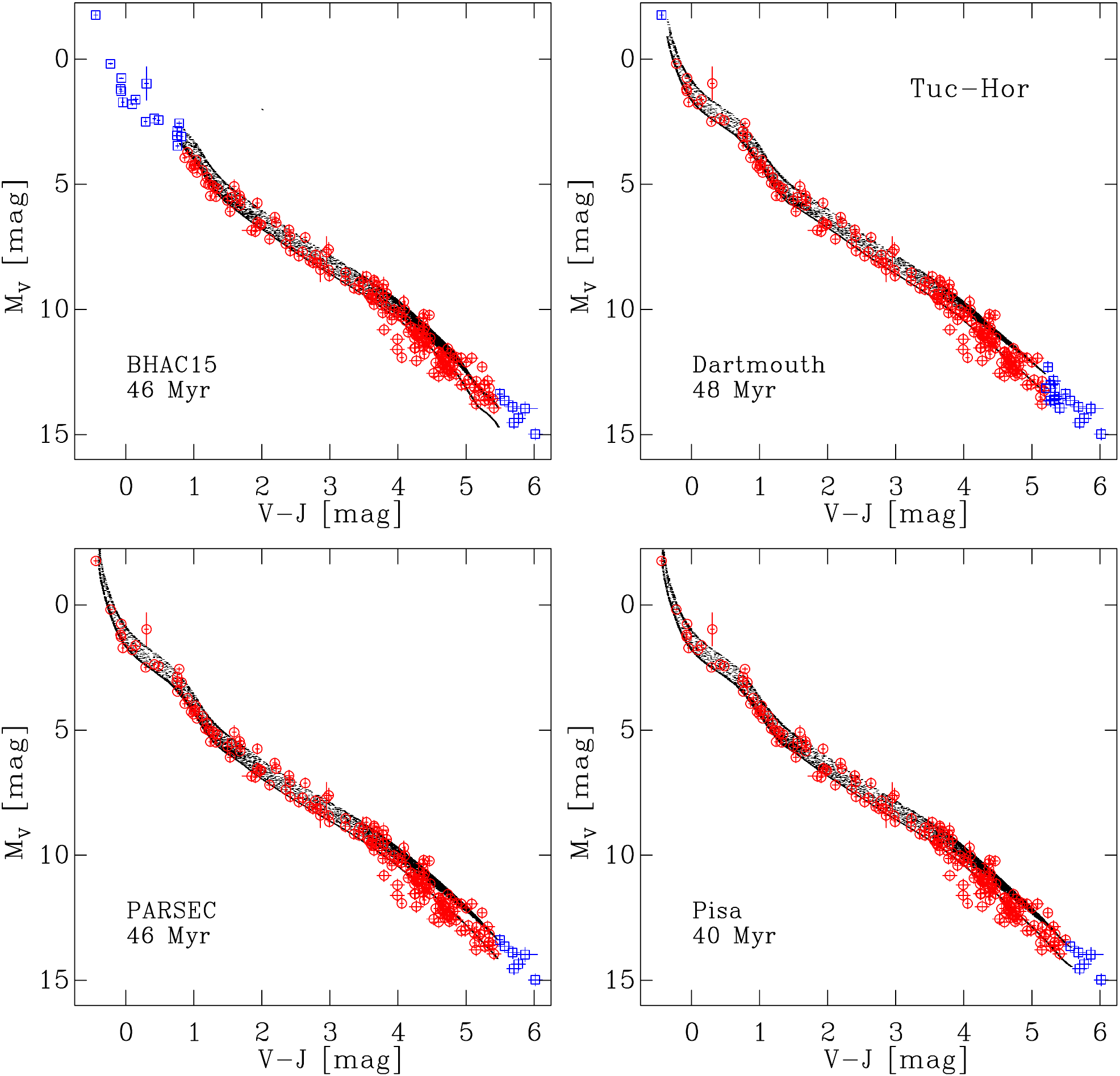} 
 \caption{Best-fitting $M_{V}, V-J$ CMDs of Tuc-Hor. The red circles represent fitted data, whereas the blue squares denote objects which are removed prior to fitting as they lie outside the area of CMD space covered by the grid of models. \textbf{Top left:} BHAC15. \textbf{Top right:} Dartmouth. \textbf{Bottom left:} PARSEC. \textbf{Bottom right:} Pisa.}
   \label{fig:tuc_hor_cmd}
\end{figure}

In essence the $\tau^{2}$ fitting statistic can be viewed as a generalisation of the $\chi^{2}$ statistic to two dimensions on the basis that both the model isochrone and photometric data are two-dimensional distributions; specifically uncertainties in colour and magnitude for the data and the widening of the model isochrone due to the effects of binarity. As in \cite[Naylor \& Jeffries (2006)]{Naylor2006}, we define the $\tau^{2}$ statistic as

\begin{equation}
\tau^{2} = -2 \sum_{i=1,\,N} \mathrm{ln}  \iint U_{i}(c-c_{i}, m-m_{i})\, \rho (c,m)\, \mathrm{d}c\, \mathrm{d}m,
\end{equation}

\noindent where $U_{i}$ represents the two-dimensional uncertainty function and $\rho$ the probability distribution of the expected model in CMD space. We create our model distributions using a Monte Carlo method to simulate $10^{6}$ stars over a given mass range and populate these using a broken power law mass function. For the inclusion of binaries we assume a uniform fraction of 50\%. Note that the best-fit age is remarkably insensitive to the adopted binary fraction. Our full grid of models range from $\mathrm{log(age)}=6.0-10.0$ in steps of $\Delta \mathrm{log(age)}=0.01\,\rm{dex}$. For each model/data combination, we then calculate the resultant $\tau^{2}$ until we find the model which minimises the above function. This is equivalent to maximising the `overlap' between the data points and the model. Fig.~\ref{fig:tuc_hor_cmd} illustrates an example of the best-fitting $M_{V}, V-J$ CMDs of Tuc-Hor, and Table~\ref{tab:tau2_results} shows the best-fit ages for each young group in our sample.

As part of our modification to the statistic, we now assign prior membership probabilities to each individual star in a given catalogue. Consequently, our best-fitting model returns posterior membership probabilities. For a given group in our sample, we expect a negligible age spread (or equivalently luminosity spread), and hence we can then use these posteriors to identify stars which appear to be non-members (based solely on CMD position) in an effort to further refine the membership lists of these young group.

\begin{table}
\begin{center}
\caption[]{Ages for the young groups in our sample. The penultimate row lists our final adopted age for each group for which the associated uncertainties represent the statistical and systematic uncertainties added in quadrature.The final row shows literature LDB ages for the BPMG and Tuc-Hor (see Section~\ref{discussion} for references) which highlights the consistency between the two age diagnostics.}
\label{tab:tau2_results}
{\scriptsize
\begin{tabular}{l c c c c c c c c c}
\hline
Model   &   \multicolumn{9}{c}{Group age (Myr)}\\
              &   AB Dor   &   Argus$^{1}$   &   BPMG   &   Carina   &   Columba   &   $\eta$ Cha   &   Tuc-Hor   &   TWA   &   32 Ori\\
\hline
BHAC15   &   $145^{+889}_{-5}$   &   $69^{+19}_{-8}$   &   $25\pm1$   &   $46^{+16}_{-3}$   &   $44^{+8}_{-3}$   &   $12\pm1$   &   $46^{+1}_{-2}$   &   $10\pm1$   &   $23^{+3}_{-2}$\\
%   &   0.38 (0.94)   &   &   &   &   &   &   &   &   \\
Dartmouth   &   $135^{+15}_{-9}$   &   $60^{+64}_{-3}$   &   $23\pm1$   &   $49^{+13}_{-5}$   &   $43^{+8}_{-3}$   &   $10\pm1$   &   $48^{+1}_{-2}$   &   $7^{+2}_{-1}$   &   $20^{+5}_{-1}$\\
%   &   0.70 (0.95)   &   &   &   &   &   &   &   &   \\
PARSEC   &   $151^{+24}_{-18}$   &   $60^{+268}$   &   $25^{+1}_{-2}$   &   $45^{+7}_{-12}$   &   $43^{+3}_{-4}$   &   $14\pm1$   &   $46\pm1$   &   $13\pm1$   &   $25^{+2}_{-5}$\\
%   &   0.56 (0.94)   &   &   &   &   &   &   &   &   \\
Pisa       &   $166^{+74}_{-23}$   &   $55^{+235}_{-5}$   &   $20^{+1}_{-2}$   &   $41^{+8}_{-6}$   &   $38\pm3$   &   $8\pm1$   &   $40^{+1}_{-2}$   &   $9\pm1$   &   $18\pm1$\\
\hline
\textbf{Adopted}   &   $\mathbf{149^{+51}_{-19}}$   &   \textbf{--}   &   \textbf{24$\pm$3}   &   $\mathbf{45^{+11}_{-7}}$   &   $\mathbf{42^{+6}_{-4}}$   &   \textbf{11$\pm$3}   &   \textbf{45$\pm$4}   &      \textbf{10$\pm$3}   &   $\mathbf{22^{+4}_{-3}}$   \\
\hline
LDB age   &   --   &   --   &   $24\pm5$   &   --   &   --   &   --   &   $40\pm3$   &   --   &   --   \\
\hline
\end{tabular}
}
\end{center}
\vspace{1mm}
\scriptsize{
{\it Notes:}\\
$^{1}$No final adopted age is given as it is unclear whether the stars in our list of members represent a single population of coeval stars (see Section~\ref{discussion}). Note also that the PARSEC models only provide an upper limit on the age of the association and therefore we do not provide a lower age uncertainty in the table.}
\end{table}

\section{Discussion}
\label{discussion}

To assess the reliability of our self-consistent, absolute isochronal age scale, we must first compare our derived ages to what are considered well-constrained ages for the same groups. Model-independent methods (such as kinematic `traceback' or `expansion' ages) should offer such a comparison, however studies have demonstrated that the ages inferred from kinematic information alone are simply too unreliable and even unreproducible when similar analyses have been performed using improved astrometric data (see e.g. \cite[Mamajek \& Bell 2014]{Mamajek2014}). LDB ages have been argued to be both \emph{accurate} and \emph{precise} to just a few Myr (see the discussion in \cite[Soderblom et al. 2014]{Soderblom2014}), and have recently been calculated for both the BPMG and Tuc-Hor\footnote[2]{Recent analyses (see e.g. Somers this volume) have demonstrated that LDB ages may be underestimated by $\simeq 10-20\%$ as a result of neglecting the effects of starspots in the evolution of pre-MS stars.}. The LDB in the BPMG has been identified by both \cite[Binks \& Jeffries (2014)]{Binks2014} and \cite[Malo et al. (2014b)]{Malo2014b}, who derive an age consistent with $24\pm5\,\rm{Myr}$. Note that our age for the BPMG is in excellent agreement with the $22\pm3\,\rm{Myr}$ derived by \cite[Mamajek \& Bell (2014)]{Mamajek2014} which was based on an analysis of the A-, F- and G-type members of the group. In addition, \cite[Kraus et al. (2014)]{Kraus2014} identified the LDB in Tuc-Hor and derived an age commensurate with $40\pm3\,\rm{Myr}$. Our isochronal ages for both of these groups are consistent with the LDB ages.

Of the remaining groups in our sample, our isochronal ages for the AB Dor moving group, Carina, Columba, $\eta$ Cha, and TWA are all consistent with previous estimates in the literature, however the advantage of our scale is the self-consistency which was previously lacking. %Note that for the AB Dor moving group our final adopted age is based only on stars with $V-J \leq 2.5\,\rm{mag}$ as the low-mass population does not provide a robust and precise age diagnostic (see the best-fit BHAC15 age in Table~\ref{tab:tau2_results}). The reason for this is that the model isochrones for ages $\gtrsim 100\,\rm{Myr}$ occupy essentially the same position in CMD space (except for very-low-mass objects with $V-J \gtrsim 4.5\,\rm{mag}$, for which the AB Dor moving group only has a few stars), resulting in a flat $\tau^{2}$ distribution over a wide range of ages. As a result of this colour cut, the adopted age does not include an age based on the BHAC15 models due to an upper mass cut-off of $1.4\,\rm{M_{\odot}}$.
The 32 Ori group has only recently begun to be investigated (see e.g. \cite[Mamajek 2007]{Mamajek2007}; \cite[Shvonski et al. 2010]{Shvonski2010}) and hence our isochronal age represents the first definitive age for this group. Finally, the reader will note that we do not provide a final adopted age for Argus, and there are two primary reasons for this. First, if Argus is indeed $\simeq 40\,\rm{Myr}$-old, then the group of 5 A-type stars (all of which were proposed as members by \cite[Zuckerman et al. 2011]{Zuckerman2011} and none of which are unresolved binaries) should be located on the zero-age main-sequence (ZAMS). The fact that none of these stars are (4 are over-luminous and 1 is under-luminous with respect to the ZAMS) suggests that these stars are not coeval, but instead represent stars at different evolutionary stages. Second, of the K- and M-type stars, less than two thirds have CMD positions commensurate with an age of $\simeq 40\,\rm{Myr}$, whereas the remainder appear significantly fainter and thus older. We are therefore of the opinion that either our list of members for Argus suffers from a high level of contamination or that the association is not physical.

\section{Summary}
\label{summary}

We present a self-consistent, absolute age scale for eight young ($\lesssim 200\,\rm{Myr}$), nearby ($\lesssim 100\,\rm{pc}$) moving groups in the solar neighbourhood. Further details concerning the methodology presented in this contribution can be found in Bell et al. (2015), but in brief our analysis involves homogeneously fitting the $M_{V}, V-J$ CMDs to derive best-fit ages using semi-empirical pre-MS models isochrones in conjunction with the $\tau^{2}$ maximum-likelihood fitting statistic. Our isochronal ages are consistent with currently available LDB ages. Such consistency instills confidence that our new isochronal age scale for young, nearby groups is robust and hence we suggest that these ages be adopted for future studies of these groups. At the moment, we are uncomfortable assigning a final, unambiguous age to Argus as it remains unclear whether our membership list for this group constitutes a single population of coeval stars.

\begin{discussion}

\discuss{G. Chabrier}{More of a (nasty) comment than a question, sorry. I'm not sure I believe your age scale given the methodology of creating the semi-empirical model isochrones you have described. First, some of the models you have adopted in your analysis have internal inconsistencies with regard to the interior physics (e.g. the stellar interior/atmospheric boundary condition) and should therefore not be adopted in the first place. Second, you derive semi-empirical $T_{\rm{eff}}$-bolometric correction relations at the age of the Pleiades and then apply these at younger ages, however there is no reason to believe that such a relation will be valid in this age regime. Third, your models rest almost entirely on a well-constrained age (and distance) for the Pleiades.}

\discuss{C. Bell}{I appreciate your concerns, however how are we to determine the ages of young stellar populations if none of the models fit the data? Given this discrepancy the \emph{only} option is to make some sort of empirical correction to the models. I'll answer your points in reverse order. First, the Pleiades does have a well-constrained age of $\simeq 130\,\rm{Myr}$ which has been derived from both the high- and low-mass population using distinct techniques and methods that rely on different aspects of stellar evolution, and so I believe the age we adopt is robust. Second, true we do derive the $T_{\rm{eff}}$-bolometric correction relation at 130\,Myr, however we attempt to account for the log$\,g$ dependence using information from the atmospheric models; specifically how the bolometric corrections vary as a function of log$\,g$. Third, again true, the majority of models do suffer from internal inconsistencies, however even if we adopt those with fully consistent treatments (such as the most recent BHAC15 models), there is still a large discrepancy between the models and the data for the Pleiades at low $T_{\rm{eff}}$. Furthermore, any internal inconsistencies are effectively `tuned out' as a result of the empirical corrections we derive and apply to the theoretical $T_{\rm{eff}}$-bolometric correction relation, and so these should not have a significant effect on the derived ages.}

\discuss{B. Zuckerman}{I'm surprised at the older ages you derive for Argus. I respect Beto Torres and collaborators enough to find their age of $\sim 40\,\rm{Myr}$ hard to reconcile with those you derive.}

\discuss{C. Bell}{I agree, it is perturbing. Interestingly, if we take the Argus members from \cite[Torres et al. (2008)]{Torres2008} and run these through the BANYAN analysis tool we find that only 11 of the 29 stars (38\%) appear to be high-probability ($\geq 90\%$) candidate members. Furthermore, of these, several have calculated kinematic distances which differ from those given in Torres et al. by more than $2-3\sigma$. Whether this is highlighting an underlying issue with the Torres et al. membership list or the BANYAN kinematic distances remains unclear.}

\discuss{J. Gagn{\'e}}{On a related note, are the members you adopt for Argus taken directly from the BANYAN analyses of Malo et al.?}

\discuss{C. Bell}{Yes, our membership list for Argus is comprised from the studies of \cite[Malo et al. (2013, 2014a,b)]{Malo2013,Malo2014a,Malo2014b}. However, we \emph{only include} candidate members if they have a measured diagnostic (e.g. radial velocity or trigonometric parallax) which is consistent with Argus membership and \emph{do not include} the other high-probability candidate members with no such measured diagnostic.}

\end{discussion}

\end{document}